\documentclass[aps,apl,preprint,superscriptaddress]{revtex4}
\usepackage{graphicx}
\usepackage{epstopdf}
\usepackage{fancyhdr}
\begin{document}

\title{Electrostatically induced phononic crystal}

\author{D. Hatanaka}
\email{hatanaka.daiki@lab.ntt.co.jp}
\affiliation{NTT Basic Research Laboratories, NTT Corporation, Atsugi-shi, Kanagawa 243-0198, Japan}

\author{A. Bachtold}
\affiliation{ICFO-Institut de Ciencies Fotoniques, The Barcelona Institute of Science and Technology, Castelldefels (Barcelona) 08860, Spain}

\author{H. Yamaguchi}
\affiliation{NTT Basic Research Laboratories, NTT Corporation, Atsugi-shi, Kanagawa 243-0198, Japan}

\begin{abstract}
The possibility of realizing an electrostatically induced phononic crystal is investigated numerically in an acoustic waveguide based on a graphene sheet that is suspended over periodically arrayed electrodes. The application of dc voltage to these electrodes exerts electrostatic force on the graphene and this results in the periodic formation of stress in the waveguide structure in a non-invasive way, unlike the cases with mass loading and air holes. This non-invasive scheme enables a bandgap, namely a $phononic$ $crystal$, to be created in the waveguide that can be used to dynamically tune the acoustic transparency in the medium. Our approach will allow the dispersion relation to be locally modified, thus modulating the temporal response of traveling acoustic phonon waves. This novel phonon architecture is promising in terms of realizing the advanced control of phonon dynamics such as waveform and dissipation engineering in the device.
\end{abstract} 

\maketitle

\hspace*{0.5em}A phononic crystal (PnC) is a promising platform for tailoring the dispersion relation of acoustic phonon waves and engineering the transmission properties \cite{narayanamurti_pnc,martinez_pnc,maldovan_nature}. Hence, the advent of this artificial phonon architecture has attracted considerable interest in phononics \cite{liu_pnc,benchabane_pnc1,mohammadi_pnc1,otsuka_pnc,boechler_pnc,laude_pnc}. The aim in the field is to develop highly functional phononic devices such as signal processors \cite{mohammadi_pnc2} and ultrahigh-$Q$ mechanical oscillators \cite{tsaturyan_pnc}, and also to open up possible applications in topological physics \cite{mousavi_topo_pnc,he_topo_pnc,painter_topo_pnc}, thermal management \cite{zen_heat,maire_heat,kodama_heat} and electro-/opto-mechanics \cite{mohammadi_phoxonic,painter_nphoton,balram_nphoton}.\\
\hspace*{1.5em}To further increase the functionality in these devices, intensive efforts have been devoted to developing dynamic PnCs that allow the temporal control of acoustic waves with an external stimulus. Most of these PnCs have been based on macroscale solid-fluid \cite{caleap_tunable_pnc,wang_tunable_pnc} or sonic crystals \cite{casadei_tunable_pnc}. In contrast, a chip-based PnC platform has recently been demonstrated that consists of a membrane resonator array with a GaAs/AlGaAs heterostructure \cite{hatanaka_pnc}. The ability to dynamically control acoustic waves by using the mechanical nonlinearity in the device has made it possible to realize functional acoustic devices \cite{hatanaka_ram,hatanaka_scirep}. However, these devices suffer from the low breakdown in the GaAs piezoelectric transduction and the narrow operation bandwidth. Additionally, in most conventional PnC devices, the formation of the bandgap still relies on their passive geometric structure, which also determines the fundamental limitation of the dynamic control.\\
\hspace*{1.5em}One of the keys to overcoming these difficulties is to use an electromechanical resonator based on graphene \cite{chen_graphene,weber_icfo1,weber_icfo2,mathew_graphene,alba_graphene} and other two-dimensional (2D) materials \cite{nicolas_wse2}. This is because these layered materials host a low mass and a large Young modulus ($m_{\rm eff}$ $\sim$ 10$^{-17}$ kg and $E$ = 1 TPa in graphene, respectively \cite{weber_icfo1}) that enhance the mode coupling nonlinearity of the resonator ($\propto$ $E/m_{\rm eff}$) \cite{eichler_cnt}. Additionally, they have extremely low rigidity that enables them to be stretched significantly by employing modest electrostatic force. This deformation induces stress without any invasive physical contact and realizes large mechanical tunability. As a result, the electrically active phonon systems using graphene and other 2D materials have demonstrated a range of dynamic control over acoustic resonant vibrations including coherent manipulation between different modes \cite{mathew_graphene,alba_graphene} and the possibility to tune the operating frequency by a large amount \cite{chen_graphene,weber_icfo1,takamura_graphene,weber_icfo2,mathew_graphene,alba_graphene}.\\
\hspace*{1.5em}In this letter, we propose an electrostatically induced PnC based on graphene electromechanical resonators, and we investigate the properties using the finite element method (FEM). The application of a dc voltage to a gate electrode array underneath a suspended graphene waveguide (WG) exerts electrostatic force on the graphene, which induces bending toward the electrodes \cite{chen_graphene,weber_icfo1,weber_icfo2,mathew_graphene,alba_graphene,nicolas_wse2}. This electrostatic stretching forms a periodic elastic potential in the WG and modulates the dispersion relation of traveling acoustic waves, enabling bandgaps to be non-invasively created. Consequently, this non-invasive approach can realize a dynamic transition in the acoustic conduction property from transparency to opacity, i.e. from WG to PnC. This also enables the temporal control of group velocity and the trapping of the acoustic waves in an arbitrary position. Thus, the graphene-based phonon architecture will open up possible applications in the fields of signal processing and electromechanics.\\
\hspace*{1.5em}The device consists of a graphene multilayer with a thickness $t=25$ $\times$ $t_{\rm g}$ = 8.5 nm ($t_{\rm g}$ is the thickness of  graphene monolayer, 0.34 nm) and density $\rho=$ $\eta$ $\times$ 2235 kg/m$^{3}$ ($\eta$ is a correction factor for contamination on a graphene surface, $\eta=$ 4.5 used in this study \cite{weber_icfo1}). The graphene sheet is suspended over a rectangular trench of length $L$ and width $w=$ 3.2 $\mu$m ($L$ $\gg$ $w$) fabricated from an insulating material such as SiO$_{\rm 2}$ as shown in Fig. 1. A gate electrode with the dimensions ($w_{\rm gx}$, $w_{\rm gy}$) = (1.60 $\mu$m, 2.88 $\mu$m) is periodically arrayed with a pitch of $a=$ 3.2 $\mu$m at the bottom of the trench and the electrodes are spatially separated in vacuum from the suspended graphene by a distance $d=$ 85 nm $\sim$ 170 nm. The geometric parameters used in this study are typical values taken from previous experimental studies on individual mechanical resonators \cite{weber_icfo1,weber_icfo2}.\\
\hspace*{1.5em}Electrostatic pressure $P_{\rm es}$ can be induced by applying a dc electric voltage to the gate electrodes and it reads
\begin{eqnarray}
P_{\rm es}=\frac{1}{2}\frac{dC_{\rm u}}{dz}V_{\rm g}^{2}=\frac{1}{2}\frac{\varepsilon_{0}}{(d-z_{\rm s})^{2}}V_{\rm g}^{2},
\end{eqnarray}
where $C_{\rm u}$, $\varepsilon_{0}$ and $z_{\rm s}$ are the capacitance per unit area formed between the graphene and the gate electrode, vacuum permittivity and the static displacement of the graphene induced by the electrostatic force, respectively. The electromechanical properties and dispersion relation are simulated in the unit structure of the WG with COMSOL Multiphysics. It should be noted that in the simulation, the graphene unit structure is given fixed boundary and Floquet periodic conditions at ($x$, $y$) = ($x$, $\pm$$w$/2) and ($\pm$$p$/2, $y$), respectively. Here, $p$ is the separation between the electrodes on which we apply the gate voltage. The electrodes that are not voltage biased are left connected to ground.\\
\hspace*{1.5em}To understand the static electromechanical property of the device, we consider a unit structure with two electrodes in unit length $p=2a$ and investigate the mechanical deformation of the graphene when we apply a gate voltage $V_{\rm g}$ to one of the two electrodes as shown in Fig. 2(a). The application of the gate voltage concentrates the electric field between the graphene and the electrode, which exerts an electrostatic force on the graphene and thus bending it toward the electrode as shown by the solid grey line in Fig. 2(b). The static displacement $z_{\rm s}$ gets larger when decreasing gate separation $d$ as shown in Fig. 2(c). The largest displacement $z_{\rm s}^{\rm max}$ of the graphene along the trench axis is between 18.6 and 7.2 nm for $d$ between 85 and 175 nm. Such large deformations cannot be achieved with conventional mechanical devices. This deformation induces large stress in the suspended graphene WG, which increases to 395 MPa at the edges when $d$ = 85 nm as shown in Fig. 2(d), and this value can also be adjusted by changing the gate voltage $V_{\rm g}$ as shown in Fig. 2(e). These results reveal that the electrostatic graphene WG devices enable the generation of periodic stress and elastic potential variations in a non-invasive way, in contrast to previous works where stress was induced by loading mass and making air holes \cite{martinez_pnc,maldovan_nature}.\\
\hspace*{1.5em}The emergence of the phononic bandgap results from the Bragg reflection of acoustic waves traveling in the periodic elastic potential. In the following, we investigate the electrostatic effect on the band dispersion in the WG. Figure 3 shows the dispersion relation in the device at various gate voltages. It exhibits parabolic dispersion curves at small gate voltages $V_{\rm g}$ = 0 V and 2V as shown in the first and second panels from the left in Fig. 3, respectively, which are a trivial property of a normal WG structure. However, an increase in the gate voltage to $V_{\rm g}=$ 4 V results in a sizeable effect of the electrostatically induced periodic stress on the dispersion curve with the formation of bandgaps as shown in the middle panel of Fig. 3. By further increasing the gate voltage to $V_{\rm g}$ = 6 V and 8 V, the spectral width of the bandgaps is broadened and the dispersion curves become flattened as shown in the second and first panels from the right in Fig. 3, respectively. Our approach can be used to dynamically manipulate the band structure and thus the acoustic wave during propagation.\\
\hspace*{1.5em}Most importantly, another advantage of this scheme is to modify the periodic pitch of the elastic potential by simply changing the configuration of the applied gate voltage. We here modify the periodic pitch $p$ between $a$, $2a$ and $3a$ (Fig. 4(a)-(c)). This engineering of the gate voltage configuration alters the periodic elastic potential and thus enables the dispersion relation to be reconfigured. With $p=a$, the gate electrode separation is short and this causes quasi-uniform deformation of the graphene sheet along the $x$-axis so that the entire WG is stretched towards the electrodes. Thus the induced stress is not strongly localized but distributed throughout the entire device as shown in Fig. 4(a). Therefore, few bandgaps emerge. On the other hand, the cut-off frequency increases greatly when increasing gate voltage due to the uniform stress as shown in the top panel of Fig. 4(d). By changing the periodic pitch to $p=2a$ and $3a$, the gate-induced deformation and stress are being localized as shown in Fig. 4(b) and (c), respectively. As a result, multiple bandgaps occur, and the cut-off frequencies are less dependent on the gate voltage, as shown in the middle and bottom panels of Fig. 4(d), respectively.\\
\hspace*{0.5cm}Changing the gate electrode pitch $a$ also modifies the bandgaps as shown in Fig. 4(e), where we used $p=a$ and $V_{\rm g}=6$ V. This clearly indicates that the spectral position of the bandgaps varies and the number of sizerable bandgaps increases when increasing pitch $a$. In particular, when $a=$ 3.2, 6.4 and 9.6 $\mu$m (dashed lines), the same gate configurations as in Figs. 4(a), 4(b) and 4(c) are realized, respectively. In this way, the electrostatic force induced by the gate electrode array allows the band dispersion in the WG to be designed with the gate voltage configuration.\\
\hspace*{1.5em}For the experimental realization of a bandgap in the PnC, it would be necessary to prepare a WG with a length $L$ longer than 100 $\times$ $p$ \cite{hatanaka_pnc}. This would correspond to $L$ between 0.32 mm and 0.96 mm in our model. Such millimeter-scale graphene can be obtained by chemical vapor deposition. The large graphene sample can be suspended by the usual dry-transfer method \cite{delft_dry_transfer} and the recently developed vacuum stack process \cite{kang_vacstack}. Although the stress that is built in during fabrication might be nonuniform along the WG, this can be compensated for by applying the appropriate $V_{\rm g}$ to each gate electrode. Thus, we expect that graphene-based PnCs can be realized with existing fabrication techniques.\\
\hspace*{1.5em}This novel PnC layout offers several advantageous features when compared to the conventional PnC devices. First, the PnC band structure is created by the electrostatic force induced by the spatially isolated electrodes. Therefore, this non-invasive approach enables on-demand tuning of the acoustic transparency of the device. Second, the gate voltage configuration determines the band structure. This not only modulates the spectral positions of the bandgaps, but it can also create a defect cavity at any desired location via local voltage modification. For instance, this can allow the strain distribution at the clamping points to be tailored in the cavity by optimizing the gate voltages, which might be used to realize a $Q$-tunable mechanical resonator \cite{tsaturyan_pnc}. Finally, the electrostatic force can also modify the mechanical nonlinearity of the device, which tunes and enhances the nonlinear parametric effect \cite{eichler_cnt}. The ability to adjust both the nonlinearity and the dispersion of the device can be useful for investigating nonlinear phenomena such as phononic solitons. This is because the soliton is generated as a result of the balance between the effects of nonlinear phase modulation and group velocity dispersion on a traveling wave \cite{kurosu_ncomm}. The control of both effects holds promise for the demonstration of the dynamic manipulation of the temporal and spectral waveforms of the acoustic wave.\\
\hspace*{1.5em}In conclusion, we have proposed a new PnC structure where the periodic stress profile can be non-invasively induced in a suspended graphene WG by applying a dc voltage to a gate electrode array. As a result the dispersion relation is modulated and bandgaps are generated. Moreover, the spectral position and width of the bandgaps can be controlled by varying the periodic arrangement of the gate voltage application. The ability to dynamically engineer the band structures opens up the possibility of developing various acoustic devices for signal processing applications, and the approach can be also used to study novel nonlinear phononic phenomena and topological physics, and to construct highly functional electromechanical circuits.\\
\\
\hspace*{1.5em}This work was supported by a MEXT Grant-in-Aid for Scientific Research on Innovative Areas “Science of hybrid quantum systems" (Grant No. JP15H05869 and JP15K21727), the Foundation Cellex, the CERCA Programme, AGAUR, Severo Ochoa (SEV-2015-0522), the grant FIS2015-69831-P of MINECO, and the Fondo Europeo de Desarrollo Regional (FEDER).\\


\newpage

\begin{figure*}[h]
\begin{center}
\vspace{-0.5cm}\hspace{-1.5cm}
\includegraphics[scale=0.65]{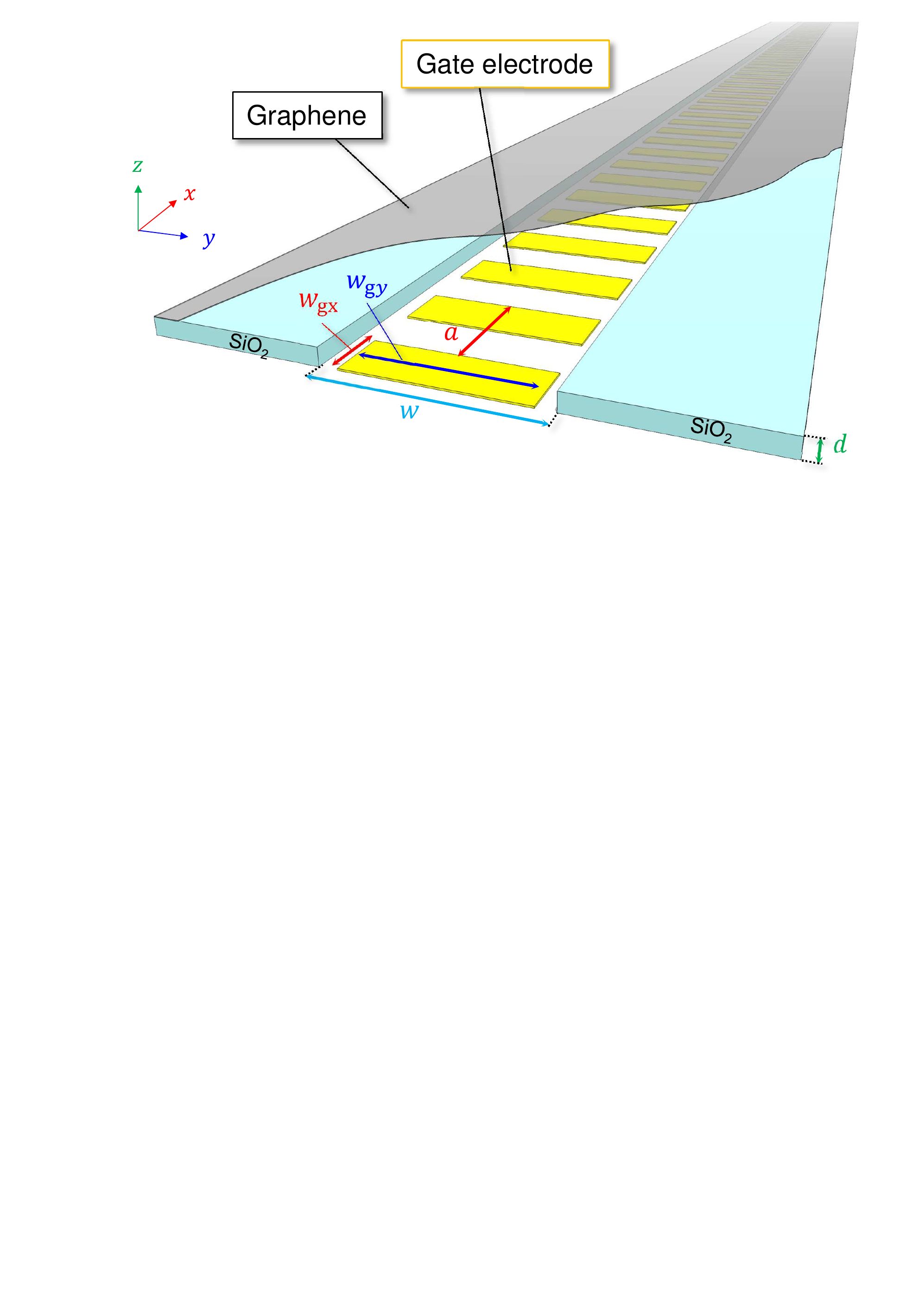}
\vspace{-11.5cm}
\caption{A schematic of a graphene-based acoustic WG where the graphene film is suspended over a trench with a gate electrode array at its bottom. Electrodes with the dimensions $w_{\rm gx}$ $\times$ $w_{\rm gy}$ are periodically arrayed with a distance of $a$ between them. Graphene with a thickness $t$ is suspended by a trench with a width $w$ and it is spatially separated from the gate electrodes by a distance $d$.}
\label{fig 1}
\vspace{-1cm}
\end{center}
\end{figure*}

\begin{figure*}[h]
\begin{center}
\vspace{-0.5cm}\hspace{-0cm}
\includegraphics[scale=0.7]{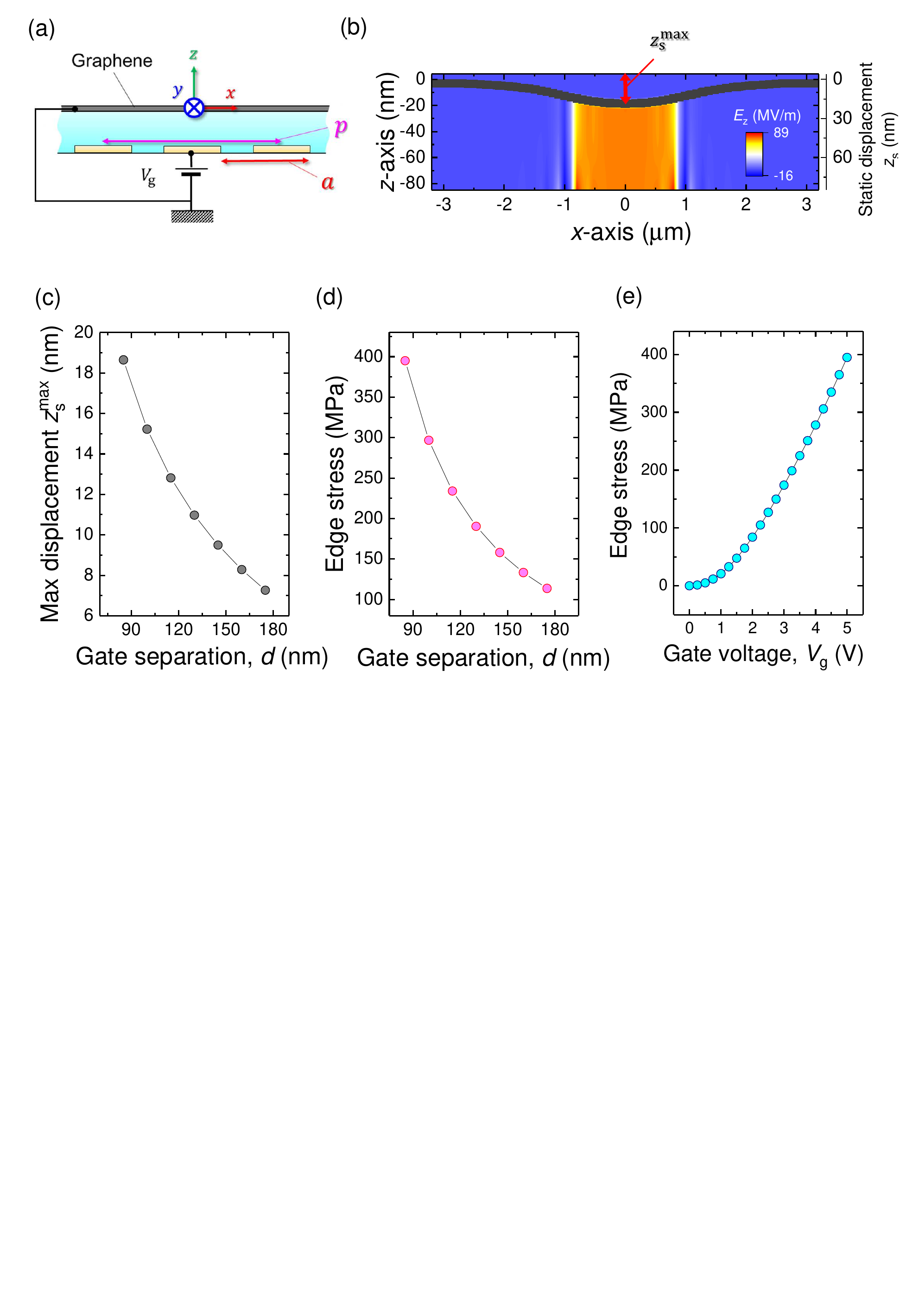}
\vspace{-9cm}
\caption{($\bf{a}$) A schematic showing the cross-section of the unit structure of the graphene WG used for FEM calculations. The unit length along the $x$ axis is $p=2a=$ 6.4 $\mu$m. Here, the dc gate voltage is applied between one electrode and the graphene. ($\bf{b}$) Electric field distribution and static displacement of the graphene sheet along the $x$-axis. The electric field in the $z$ direction $E_{\rm z}$ is represented on the $x$-$z$ plane at $y$ = 0 when applying $V_{\rm g}$ = 5 V. The field $E_{\rm z}$ is strongly confined in a vacuum gap with $d=$ 85 nm between the electrode and graphene. This exerts electrostatic force on the graphene and bends it toward the electrode (grey line). ($\bf{c}$) and ($\bf{d}$) The gate-graphene separation $d$ dependence of the maximum static displacement $z_{\rm s}^{\rm max}$ (see (b)) and the stress at the clamping points ($x$, $y$) = (0, $\pm$ $w/2$), which are electrostatically induced by $V_{\rm g}=$ 5 V. ($\bf{e}$) The induced stress as a function of $V_{\rm g}$ at $d$ = 85 nm.}
\label{fig 2}
\vspace{-0.5cm}
\end{center}
\end{figure*}

\begin{figure*}[h]
\begin{center}
\vspace{-0.5cm}\hspace{-3cm}
\includegraphics[scale=0.85]{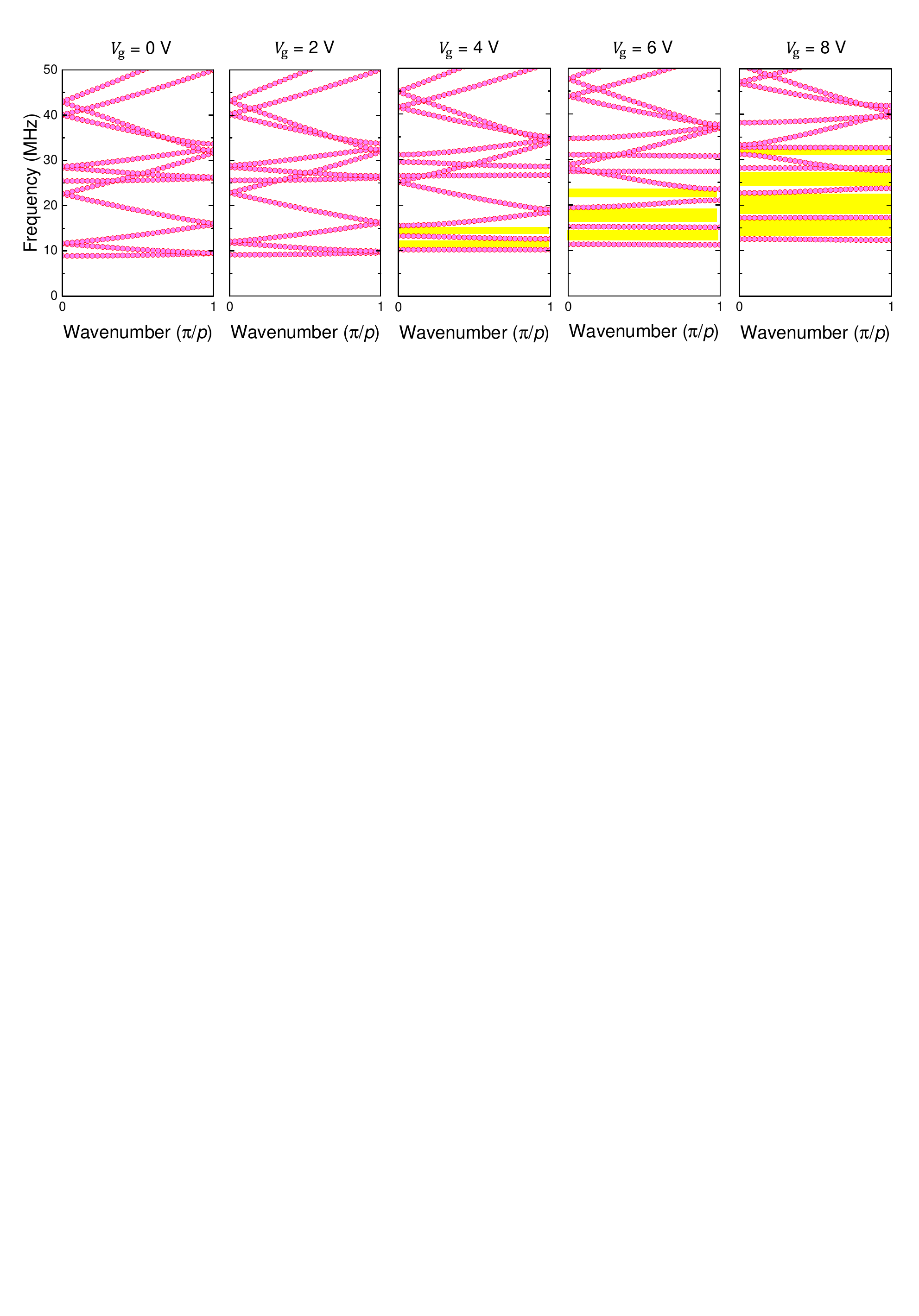}
\vspace{-17.5cm}
\caption{The dispersion relation of the graphene-based WG with $d=$ 130 nm at various gate voltages $V_{\rm g}$s. The dispersion is simulated by applying the Floquet periodic condition to the unit structure shown in Fig. 2(a). An increase in $V_{\rm g}$ creates the bandgaps (yellow) and the reduction in the dispersion slopes (pink circles).}
\label{fig 3}
\vspace{-0.5cm}
\end{center}
\end{figure*}

\begin{figure*}[p]
\begin{center}
\vspace{-2.5cm}\hspace{-3cm}
\includegraphics[scale=0.68]{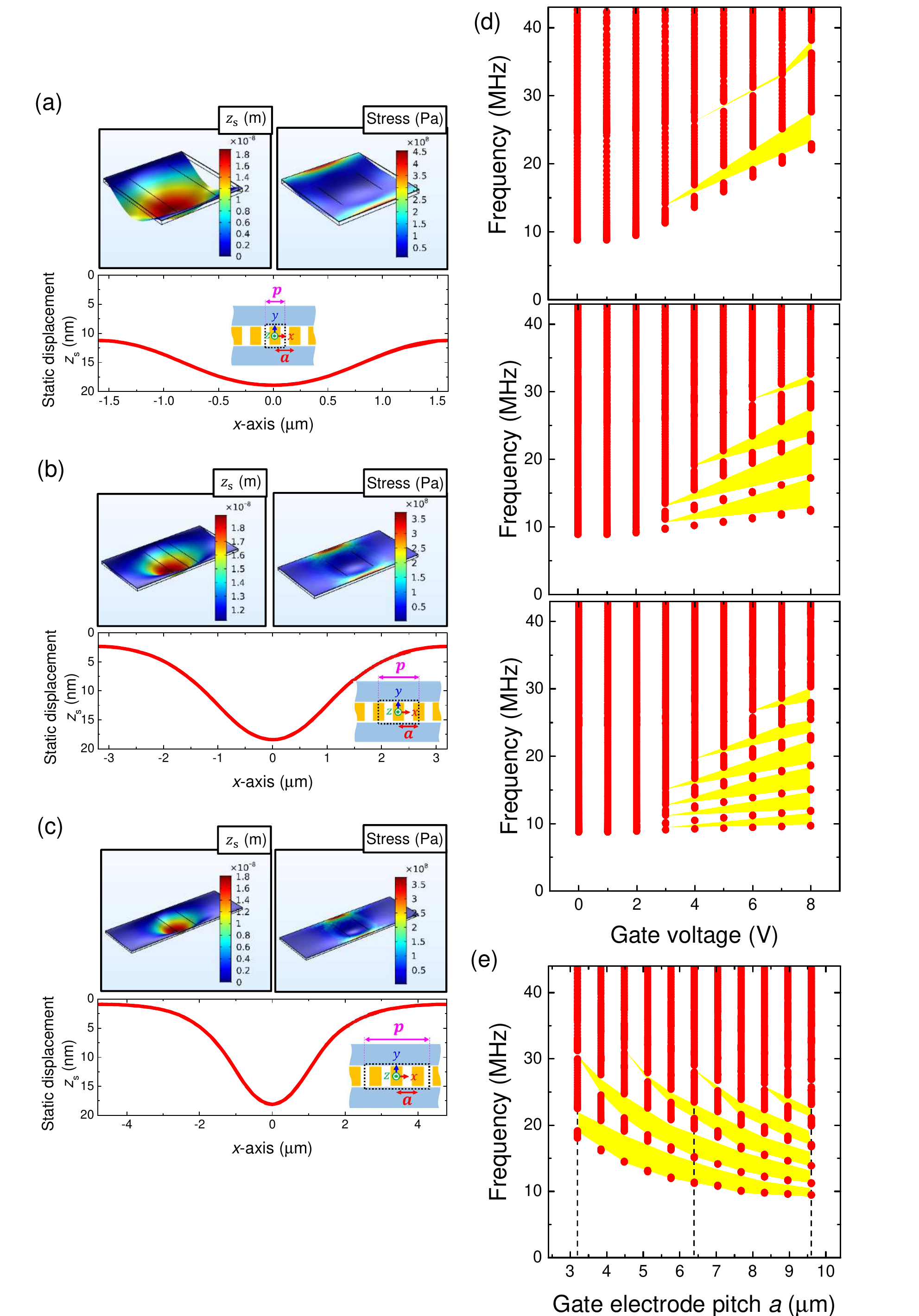}
\vspace{-0.2cm}
\caption{($\bf{a}$)-($\bf{c}$) The simulated profile of the static displacement $z_{\rm s}$ (left panel) and stress (right panel) in the unit structure of the graphene WG with $d=$ 130 nm and $p=a, 2a$ and 3$a$ when applying $V_{\rm g}=$ 8 V. The static displacement $z_{\rm s}$ in the $x$-$z$ plane at $y=$ 0 as shown in the bottom panel. The inset in the bottom panel is a schematic of the cross-section and the $x$-$y$-$z$ coordinate. The black dotted frame denotes the unit structure with length $p$. ($\bf{d}$) The eigenfrequencies as a function of the gate voltage $V_{\rm g}$ with $p=a, 2a$ and $3a$ as shown in the top, middle and bottom panels, respectively. The bandgaps are highlighted in yellow. ($\bf{e}$) The periodic pitch $a$ dependence of the eigenfrequencies in the device at $V_{\rm g}=$ 6 V. The bandgap is highlighted in yellow. The black dashed lines denote $a=$ 3.2, 6.4 and 9.6 $\mu$m where the same configurations are realized as in (a), (b) and (c), respectively.}
\label{fig 4}
\vspace{-0.5cm}
\end{center}
\end{figure*}

\end{document}